\documentclass{article}
\usepackage{arxiv}
\usepackage[utf8]{inputenc} % allow utf-8 input
\usepackage[T1]{fontenc}    % use 8-bit T1 fonts
\usepackage{hyperref}       % hyperlinks
\usepackage{url}            % simple URL typesetting
\usepackage{booktabs}       % professional-quality tables
\usepackage{amsmath,amsfonts,amssymb,amsthm}       % blackboard math symbols
\usepackage{nicefrac}       % compact symbols for 1/2, etc.
\usepackage{microtype}      % microtypography
\usepackage{lipsum}
\usepackage{graphicx}
\usepackage{lscape}
\graphicspath{ {./images/} }

\title{Modeling Treatment Effect Modification in Multidrug-Resistant Tuberculosis in an Individual Patient Data Meta-Analysis}

\author{Yan Liu$^1$, Mireille Schnitzer$^{2,3}$, Guanbo Wang$^{1}$, Edward Kennedy$^{4}$,\\
\textbf{Piret Viiklepp$^{5}$, Mario H. Vargas$^{6}$, Giovanni Sotgiu$^{7}$, Dick Menzies$^{8,9}$ and }\\
\textbf{Andrea Benedetti$^{1,8,9}$}

\vspace{8mm}\\
$^{1}$ \normalsize{\textit{Department of Epidemiology, Biostatistics and Occupational Health,}}\\ \normalsize{\textit{McGill University, Montréal, Québec, Canada}}\\
$^{2}$ \normalsize{\textit{Faculty of Pharmacy, Université de Montréal, Montréal, Québec, Canada}}\\
$^{3}$ \normalsize{\textit{Department of Social and Preventive Medicine,Université de Montréal, Montréal, Québec, Canada}}\\
$^{4}$ \normalsize{\textit{Department of Statistics \& Data Science, Carnegie Mellon University, Pittsburgh, USA}}\\
$^{5}$ \normalsize{\textit{National Institute for Health Development, Estonia}}\\
$^{6}$ \normalsize{\textit{Instituto Nacional de Enfermedades Respiratorias, Mexico City, Mexico.}}\\
$^{7}$ \normalsize{\textit{Clinical Epidemiology and Medical Statistics Unit, Department of Medical,}}\\ \normalsize{\textit{Surgical and Experimental Sciences, University of Sassari, Sassari, Italy}}\\
$^{8}$ \normalsize{\textit{Respiratory Epidemiology and Clinical Research Unit, McGill University Health Centre,}}\\\normalsize{\textit{ Montréal, Québec, Canada}}\\
$^{9}$ \normalsize{\textit{Department of Medicine, McGill University, Montréal, Québec, Canada}} 
}
\begin{document}
\maketitle

\begin{abstract}
Effect modification occurs while the effect of the treatment is not homogeneous across the different strata of patient characteristics. When the effect of treatment may vary from individual to individual, precision medicine can be improved by identifying patient covariates to estimate the size and direction of the effect at the individual level. However, this task is statistically challenging and typically requires large amounts of data. Investigators may be interested in using the individual patient data (IPD) from multiple studies to estimate these treatment effect models. Our data arise from a systematic review of observational studies contrasting different treatments for multidrug-resistant tuberculosis (MDR-TB), where multiple antimicrobial agents are taken concurrently to cure the infection. We propose a marginal structural model (MSM) for effect modification by different patient characteristics and co-medications in a meta-analysis of observational IPD. We develop, evaluate, and apply a targeted maximum likelihood estimator (TMLE) for the doubly robust estimation of the parameters of the proposed MSM in this context. In particular, we allow for differential availability of treatments across studies, measured confounding within and across studies, and random effects by study.
\end{abstract}


\section{Introduction}
Multidrug-resistant tuberculosis (MDR-TB), a form of tuberculosis (TB) with high mortality, is caused by bacteria resistant to at least the two most effective anti-TB drugs, isoniazid and rifampicin. According to the World Health Organization (WHO) (2020), a global total of 206 030 patients with MDR-TB or rifampicin-resistant(RR) TB were detected and notified of their infection in 2019, a $10\%$ increase compared to cases in 2018. However, the latest data reported to WHO show a treatment success rate for MDR/RR-TB of $57\%$ globally.\cite{WHO}
Treating MDR-TB is challenging as a result of the heterogeneity of patients' characteristics (age, sex, HIV or other comorbidities), disease characteristics (extent and prior treatment), mycobacteria itself (different patterns of additional resistance) and characteristics of drugs (more toxicity and less effect of second-line drugs). \cite{Muller, Isaakidis} Patients are typically prescribed a combination of four or more antimicrobial agents depending on the therapeutic phase and drug resistance pattern, if known.\cite{WHO_treat} In addition, the effect of a treatment regimen may vary by an individual's characteristics and the specific combination of medications. When the effect and drug resistance pattern may vary from individual to individual, precision medicine can be improved by identifying patient covariates to estimate the size and direction of the effect at the individual level. In other words, identifying effect modifiers and assessing effect modification between different patient subgroups should be considered. 
However, this task is statistically challenging and typically requires large amounts of data so that treatment effects may be well-estimated for different combinations of covariate values. One may impose a working model in order to smooth (or summarize) the covariate-specific effects rather than estimate a separate effect for each possible combination of patient covariates.\cite{Peterson,Powers,Asma} When working with observational data one must also adjust for all potential confounders of the treatment-outcome relationship, which can be accomplished via propensity scores and/or outcome regression modeling.\cite{Robins_Hernan} One way to model effect modification in a simple binary treatment setting is through a marginal structural model (MSM) for the conditional average treatment effect (CATE).\cite{Asma, Zhao} This model may be interpreted as the relationship between covariates and the expected treatment effect where the treatment effect is defined through a contrast of counterfactual outcomes. In non-meta-analytical settings, doubly robust estimators have been proposed for the estimation of a parametric MSM for the CATE \cite{Asma} as well as for nonparametric CATE models. \cite{Edward} 

Due to the large data requirements for estimating effects across patient subgroups, investigators may be interested in using individual participant data from multiple studies to fit these treatment effect models. In our study, the data were extracted from 31 observational studies~\cite{Ahuja} which contrasted different treatment regimens for patients with MDR-TB, where multiple antimicrobial agents are taken concurrently by a patient over a long period. Our objective is to perform an individual participant data (IPD) meta-analysis~\cite{Mills} to investigate the impact of different patient and treatment characteristics on the average treatment effect (ATE) of 15 anti-TB medications. 

In this project, we propose a targeted maximum likelihood estimator (TMLE)~\cite{van&Rose} for the estimation of the parameters of the CATE MSM, in the  meta-analytical context. For estimation of ATE, TMLE depends on two components: an outcome regression conditional on treatment and covariates; and, weights comprised of the inverse of the propensity score where the propensity score is the probability of treatment conditional on covariates.\cite{van2014}
TMLEs are plug in estimators with asymptotic properties that use a targeting step to optimize the bias-variance trade-off for the target parameter.  In our setting, the estimator we propose allows for differential availability of treatments across studies and random effects by study due to measured and unmeasured characteristics of the study-specific populations. 

In Section \ref{Pooled Observational Studies of MDR-TB}, we describe the MDR-TB data structure and our parameters of interest. Section \ref{Models and Algorithms} introduces the TMLE procedure. We also describe a clustered influence function-based variance estimator. In Section \ref{Simulation Study}, we present the results of simulation studies to demonstrate the properties of the estimator under different scenarios. Then in Section \ref{MDR-TB Data Analysis}, we provide the results based on TMLE analysis of effect modification for 15 anti-TB medications using the combined IPD of the 31 observational studies.

\section{Pooled Observational Studies of MDR-TB}
\label{Pooled Observational Studies of MDR-TB}
The application data consist of IPD derived from 31 observational studies resulting in a total of 9290 MDR-TB patients. The updated systematic review [Ahuja] extended three previous systematic reviews.\cite{johnston, orenstein, akcakir} We use a one-stage approach where we pool all IPD across studies to analyze them in a model accounting for clustering and random effects. 
This IPD were collected from cohorts of adults and the year of studies ranged from 1995 to 2009. The information collected for each patient includes demographics (age and sex), past TB history, clinical characteristics (pre-treatment sputum smear results for acid-fast bacilli (AFB) and culture, chest radiography, HIV infection, drug susceptibility test (DST) results), anti-microbial medications given, and outcomes.

\subsection{Data Structure}
\label{Data Structure}
\subsubsection{Outcome}
In the pooled dataset, the binary outcome $Y$ represents the treatment success (the treatment was completed and cured the disease) versus treatment failure (the patient was still culture positive for MDR-TB, experienced a relapse or died).\cite{falzon2011} A patient's outcome realization is defined as lowercase $y_{ij}$ where ${(i,j)}$ refers to patient $i\in C_j$ in study $j$ and $C_j$ be the set of indices of patients in study $j$ where $j\in (1,2,\cdots,J)$. In the MDR-TB data, there are pooled data from 31 studies, i.e. $J=31$. 
\subsubsection{Treatments and Treatment Availabilities}
There are 15 antimicrobial agents observed in the data: ethambutol (EMB), ethionamide (ETO), ofloxacin (OFX), pyrazinamide (PZA), kanamycin (KMA), cycloserine (CS), capreomycin (CAP), para-aminosalicylic acid (PAS), prothionamide (PTO), streptomycin (SM), ciprofloxacin (CIP), amikacin (AMK),  later-generation fluoroquinolones (LgFQ), rifabutin (RIF) and group five level drugs (Gp5). LgFQ included levofloxacin, moxifloxacin, gatifloxacin and sparfloxacin.\cite{Ahuja} Gp5 comprised of  amoxicillin-clavulanate, macrolides (azithromycin, roxithromycin, and clarithromycin), clofazimine, thiacetazone, imipenem, linezolid, high dose isoniazid, and thioridazine.\cite{Ahuja} We use $k$ to index the $k$-th antimicrobial agent, where $k=1,2,\dots,15$ in this case.  The binary random variable $A^{(k)}$ indicates the exposure to medication $k$ with patient realizations $a_{ij}^{(k)}$. 

Not all treatments are observed in each study; we assume that a given treatment was available for a given study's participants based on whether we have observed that this treatment was taken by any patient in the given study.\cite{Guanbo} The binary variable $D^{(k)}$ is defined as the treatment availability of treatment $k$ in a given study. Specifically, $d_{ij}^{(k)}=d_{j}^{(k)}=1$ means that the treatment $k$ is available to any subject $i$ in study $j$, and is true if at least one patient in this study was prescribed the treatment $k$, and otherwise $d_{ij}^{(k)}=0$.

\subsubsection{Baseline Covariates and Resistance Information}
The baseline covariates consist of two study level covariates $\boldsymbol{S}$ (the start year of MDR-TB treatment and the income group of the country of the study) and six individual level covariates $\boldsymbol{W}$ (age, sex, AFB results, HIV infection, cavitation status on chest radiography and past TB history). 

Resistance information based on DST is defined as the binary variable $R^{(k)}$. In this dataset, drug resistance information is available for eight medications. Thus, we denote $r^{(k)}=1$ if the patient was found to be resistant to the treatment $k$ and otherwise $r^{(k)}=0$ which includes the situations where the patient's infection was susceptible to the treatment or was not known to be resistant to this treatment.

\subsubsection{Observed Data Structure}
The observed data can be written as
$\boldsymbol{O}=[\boldsymbol{S}, \boldsymbol{W}, \{A^{(k)}, D^{(k)}, R^{(k)}; k=1,2,\dots,15\},Y]$. 
We will rewrite $\boldsymbol{R}=\{R^{(k)};k=1,\cdots,15\}$, similarly for  $\boldsymbol{D}$ and $\boldsymbol{A}$. Then the data structure is $\boldsymbol{O}=(\boldsymbol{S}, \boldsymbol{D}, \boldsymbol{W}, \boldsymbol{R}, \boldsymbol{A}, Y)$.

\subsubsection{Definition of Counterfactual Notation}
In this data, there is differential availability of treatments across studies. 
In order to define a generalized parameter of interest, we define counterfactual notation under the availability of a given treatment.\cite{Guanbo} We define counterfactual exposure to treatment $A^{(k)}\{d^{(k)}=1\}$ as the patient's counterfactual usage of treatment $k$ had the patient had access. Then we define the counterfactual outcome $Y\{d^{(k)}=1,a^{(k)}\}=Y\{a^{(k)}\}$ as the patient's outcome that would have occurred had treatment $k$ been available and taken.

\subsection{Parameter of Interest and Assumptions}
\label{Parameter of Interest and Assumptions}
As in previous work, we aim to estimate the parameters of interest defined with respect to a global population which refers to the union of super-populations specific to each study in this dataset.\cite{Mireille2015, Guanbo}
We define a non-parametric structural equation model (NPSEM), which assumes a time ordered data generating structure, in the Supplementary Materials Appendix A. Then the counterfactual likelihood is derived under NPSEM and the parameter of interest is defined with respect to this likelihood.

In this project, for a given medication $k$, we are interested in estimating the coefficients of potential effect modifiers $\boldsymbol{V}^{(k)}$ of the effect of medication $k$ in an MSM for the CATE.\cite{Asma} To define the corresponding parameters, we recall that patients may take multiple medications concurrently. Given a medication $k$, we conduct the analysis by treating all other medications as confounders. For ease of notation, we define the adjustment set as $$\boldsymbol{X}^{(k)}=[\boldsymbol{S}, \boldsymbol{W}, 
R,\{D^{(k^*)}, A^{(k^*)}; for~\forall k^* \in (1,\cdots,15)~s.t.~k^*\neq k\} ]$$
with realizations $x_{ij}^{(k)}$ for individual patients. Here, the symbol asterisk indicates treatments other than the treatment of interest.
The set of potential effect modifiers investigated is a subset of the adjustment set $\boldsymbol{X}^{(k)}$.~\cite{Asma, Zhao} The random covariate vector is $\boldsymbol{V}^{(k)}=\{1,V^{(k)}_1,...,V^{(k)}_p\}$. Then we can model the CATE of treatment $k$, denoted $\psi\{\boldsymbol{V}^{(k)}\}$, as a linear function of potential effect modifiers such that
$$\mathbb{E}[Y\{a^{(k)}=1\}-Y\{a^{(k)}=0\}|\boldsymbol{V}^{(k)}]=
    \psi\{\boldsymbol{V}^{(k)};\boldsymbol{\beta_{V}}^{(k)}\}=\{\boldsymbol{V}^{(k)}\}^\intercal\boldsymbol{\beta_{V}}^{(k)}$$
where the symbol $^\intercal$ indicates a transpose and  $\boldsymbol{\beta_{V}}^{(k)}=\{\beta^{(k)}_0,\beta^{(k)}_1,...,\beta^{(k)}_p\}\in \mathbb{R}^{p+1}$. The parameter  $\beta^{(k)}_m, m\neq 0$ represents the difference in the expected causal effect related to a single unit change of the effect modifier $V_m^{(k)}$ and $\beta^{(k)}_0$ is the intercept term. 

Identifying the parameter of interest requires some assumptions that allow us to write the parameter in terms of distributions of the observed data.\cite{Guanbo}

(a) Consistency: The counterfactual outcome $Y\{d^{(k)}=1,a^{(k)}=1\}$, where treatment $k$ was available to and taken by the patient, is the same as the observed outcome for patients who in fact had access to the treatment and took it. Additionally, the counterfactual outcome had the patient not taken this treatment $Y\{a^{(k)}=0\}$ is equal to the observed outcome for those who had not taken the treatment, either due to unavailability or other reasons. In particular, the first consistency assumption may fail if this treatment was only taken temporarily by the patient prior to a more effective antimicrobial being substituted in.

(b) Positivity: For the CATE to be estimable without extrapolation, we also need positivity assumptions. Specifically, the conditional probability of being treated for each patient given the drug's availability $Pr[A^{(k)}\{d^{(k)}=1\}=1|\boldsymbol{X}^{(k)}=\boldsymbol{x}^{(k)},D^{(k)}=1]$ and the conditional probability of not being treated $Pr\{A^{(k)}=0|\boldsymbol{X}^{(k)}=\boldsymbol{x}^{(k)}\}$ must both be positive. This may fail if contraindications exist in the covariates $\boldsymbol{X}^{(k)}$, rendering treatment with $A^{(k)}$ impossible. Furthermore, $Pr\{D^{(k)}=1|\boldsymbol{S}^{(k)}=\boldsymbol{s}^{(k)}\}$, the probability of availability of treatment $k$ conditional on the study-level covariates must also be positive. We also assume that the probability of treatment availability conditional on all measured covariates is only a function of the study-level covariates. Positivity is violated when, for example, certain studies occurred in a time or country where some drugs were not on the market, but only if time or country is a study-level confounder. 

(c) Transportability: The counterfactual outcomes had the patient had access to the treatment and taken it are independent of treatment availability conditional on measured covariates, i.e. $Y^{(k)}\{d^{(k)}=1, a^{(k)}=1\}\perp D^{(k)}|\boldsymbol{X}^{(k)}$. This means that we can use the measured covariates to fit models where the treatment is available and use those model fits to estimate overall effects. 

(d) In addition, we require unconfoundedness: that the counterfactual outcomes had the patient taken this treatment be independent of the treatment assignment conditional on the measured covariates and treatment availability. i.e. $Y^{(k)}\{a^{(k)}=1\}\perp A^{(k)}|D,\boldsymbol{X}^{(k)}$. Moreover, the counterfactual outcomes had the patient not taken this treatment, $Y^{(k)}\{a^{(k)}=0\}$, are independent of the treatment assignment given the measured covariates.

Under these assumptions, the CATE can be written as:
\begin{align*}
  \psi\{\boldsymbol{V}^{(k)}\}
=&\mathbb{E}[\mathbb{E}\{Y|A^{(k)}=1,\boldsymbol{X}^{(k)}\}|\boldsymbol{V}^{(k)}]-\mathbb{E}[\mathbb{E}\{Y|A^{(k)}=0,\boldsymbol{X}^{(k)}\}|\boldsymbol{V}^{(k)}] 
\end{align*}
Both right-hand terms can be estimated from the observed data and thus the CATE is nonparametrically identifiable. A proof is provided in Supplementary Materials Appendix B.

\section{Models and Algorithms}
\label{Models and Algorithms}
\subsection{Outcome and Propensity Score Models }
\label{Outcome and Propensity Score Models }
  For ease of notation, we will drop the notation $k$ for this section, and consider $A=A^{(k)}$ for a given $k$ with $\boldsymbol{X}=\boldsymbol{X}^{(k)}$ the adjustment set for $k$ as previously defined. 
 The doubly robust estimators presented in this section require the estimation of two quantities, the conditional outcome expectation and propensity score. We define the former as $\overline{Q}(A,\boldsymbol{X})=Pr\{Y(A=a)=1|\boldsymbol{X}\}$, the probability of counterfactual treatment success conditional on the baseline covariates $\boldsymbol{X}$. The propensity score is $g(A|\boldsymbol{X})=Pr(A=a|\boldsymbol{X})$, i.e. the conditional probability of treatment given $\boldsymbol{X}$. 
 
 In our setting with treatment availability variable $D$ and the assumptions listed in Section~\ref{Parameter of Interest and Assumptions}, we note that $$\overline{Q}(1,\boldsymbol{X})=Pr(Y=1|A=1,\boldsymbol{X})=Pr(Y=1|D=1,A=1,\boldsymbol{X}).$$ 
 Thus, we can estimate this quantity by fitting a regression model using the subgroup of subjects who received this treatment, which is necessarily a subset of those who had access to the treatment. 
On the other hand, $\overline{Q}(0,\boldsymbol{X})=Pr(Y=1|A=0,\boldsymbol{X})$ includes both patients who did not have access to the given treatment and patients who did but did not receive this treatment.

For the propensity score, the probability of being treated for each patient is:
 $$g(1|\boldsymbol{X})
  =\underbrace{Pr(A=1|D=1,\boldsymbol{X})}_{g_{1}}\underbrace{Pr(D=1|\boldsymbol{S})}_{g_{2}}.$$
The first component ${g_{1}}$ may be estimated by fitting a model using patients who had access to the treatment. The second part ${g_{2}}$ can be obtained by regressing treatment availability on study level covariates $\boldsymbol{S}$. This second regression takes the study as the unit. Then, we can write the probability of not being treated as $g(0|\boldsymbol{X})=1-g(1|\boldsymbol{X})=1-g_1\cdot g_2$.

\subsection{Efficient Influence Function}
\label{Efficient Influence Function}
The efficient influence function (EIF) for a specific parameter is the influence function that achieves the efficiency in the given space of semi-parametric models.\cite{semipara2006} The EIF defines the linear approximation of any efficient and regular asymptotically linear estimator. The EIF  for the coefficients $\boldsymbol{\beta_{\boldsymbol{V}}}$ in a working i.i.d. model space $\mathcal{M}$ is $$\mathcal{D}_{\boldsymbol{\beta_{V}}}= M^{-1}\mathcal{D}$$ where
\begin{align}
    \mathcal{D}=\biggl[\biggl\{\frac{I_{A=1}}{g(1|\boldsymbol{X})}-\frac{I_{A=0}}{g(0|\boldsymbol{X})}\biggr\}\{Y-\overline{Q}(A,\boldsymbol{X})\}+\overline{Q}(1,\boldsymbol{X})-\overline{Q}(0,\boldsymbol{X})-\boldsymbol{V}^\intercal\boldsymbol{\beta_V}\biggr]\boldsymbol{V}
\end{align}
with normalizing matrix $M=-\mathbb{E}(\frac{\partial \mathcal{D}}{\partial \boldsymbol{\beta_V}})$. 
A proof is provided in Supplementary Materials Appendix C. An equivalent result was also given by Rosenblum and van der Laan. \cite{VanRose2010}

\subsection{TMLE}
\label{TMLE}
The general TMLE procedure was proposed by van der Laan and Rubin.\cite{van.TMLE} Our proposed procedure takes estimates of the conditional expected outcome  $\overline{Q}(a,\boldsymbol{X})$ and updates them using information from the propensity score. 

The first step is to produce initial estimates for $\overline{Q}(a,\boldsymbol{X})$ for $a=1$ and 0, denoted as $\overline{Q}_n(1,\boldsymbol{X})$ and $\overline{Q}_n(0,\boldsymbol{X})$, respectively. For each of $a=1$ and 0, we run a weighted logistic regression of Y with offset $logit\{\overline{Q}_n(a,\boldsymbol{X})\}$ and covariates corresponding to the set of potential effect modifiers. The weights are 
$A/{g_n(1|\boldsymbol{X})}$ for $a=1$ and $(1-A)/{g_n(0|\boldsymbol{X})}$ for $a=0$, respectively. We then set the updated $\overline{Q}^{*}_n(a,\boldsymbol{X})$ equal to the predicted values from the above logistic regression. Finally, we fit a linear regression of $\overline{Q}^{*}_n(1,\boldsymbol{X})-\overline{Q}^{*}_n(0,\boldsymbol{X})$ on the potential effect modifiers in order to obtain the TMLE estimates, $\hat{\boldsymbol \beta}^{TMLE}_{\boldsymbol V}$, of the parameters of interest.

We note that this TMLE solves the equation $\sum _{j=1}^J \sum _{i\in C_j} \mathcal{D}_{ij,n}(\hat{\boldsymbol \beta}^{TMLE}_{\boldsymbol V})=0$. The estimator thus has the properties of double robustness and local efficiency. We also describe a closely related augmented inverse probability of treatment weighted estimator (A-IPTW)~\cite{Scharfstei} in Supplementary Materials Appendix D. In finite samples, TMLE has been shown to perform better than A-IPTW if the positivity assumption is nearly violated or the true values of the parameters of interest are close to the parameter space boundaries.\cite{Porter}

\subsection{Influence Function-based Variance Estimator}
\label{Influence Function-based Variance Estimator}
Standard errors can be estimated for the TMLE using a large-sample sandwich estimator of the efficient influence function under consistency of  $\overline{Q}_n(a,\boldsymbol X)$ and $g_n(a|\boldsymbol X)$.\cite{semipara2006} Let $\boldsymbol{\beta_V}_0$ be the true value of $\boldsymbol{\beta_V}$ and $\boldsymbol{\hat{\beta}_V}$ be the TMLE estimate. 

Under regularity conditions, we can write the linear approximation of the estimator as \cite{Guanbo}
 $$\sqrt{n}(\boldsymbol{\hat{\beta}_V}-\boldsymbol{\beta_V}_0)\approx \frac{1}{\sqrt{n}}\sum _{j=1}^J \sum _{i\in C_j}\mathcal{D}_{{ij}}(\boldsymbol{\beta_{V}}_0)$$
 where $\mathcal{D}_{{ij}}(\boldsymbol{\beta_{V}}_0)$ is the influence function at the true values of $\overline{Q}(a,\boldsymbol X_{ij})$ and $g(a|\boldsymbol X_{ij})$ for each subject.

 In order to estimate the variance of the parameter of interest while taking clustering by studies into account, we only assume independence between studies, and not individuals within the same study. Within study $j$,  we denote the $(p+1)\times (p+1)$ dimension variance-covariance matrix of the efficient influence function as $\boldsymbol{\sigma}^2_j$. We denote the $(p+1)\times (p+1)$ dimension variance-covariance matrix of the efficient influence function of any two different subjects in study $j$ as $\boldsymbol{\rho}_j$. Both of these quantities can be estimated from the observed data. Then, for large $J$, the variance-covariance matrix of $\hat{\boldsymbol \beta}_{\boldsymbol V}$ can be estimated using:~\cite{Mireille2014}
\begin{align*}
Var(\boldsymbol{\hat{\beta}_V})\approx &\frac{1}{n^2}diag\biggl\{\sum _{j=1}^J
\biggl( \sum _{i,m\in C_j,i\neq m}E[\mathcal{D}_{ij,n}({\hat{\boldsymbol\beta}_{\boldsymbol V}}) \{\mathcal{D}_{mj,n}({\hat{\boldsymbol\beta}_{\boldsymbol V}})\}^\intercal]\\
 &\ \ \ + \ \sum _{i\in C_j}E[\mathcal{D}_{ij,n}({\hat{\boldsymbol\beta}_{\boldsymbol V}})\{\mathcal{D}_{ij,n}({\hat{\boldsymbol\beta}_{\boldsymbol V}})\}^\intercal]\biggr)
\biggr\}\\
= &\frac{1}{n^2}diag\biggl[\sum _{j=1}^J\biggl\{ n_j(n_j-1)\boldsymbol{\rho}_j+n_j\boldsymbol{\sigma}^2_j\biggr\}\biggr]
\end{align*}
where $n_j$ is the size of study $C_j$. This variance estimator is only valid for larger numbers of clusters and  consistent estimates of both the outcome and the propensity score.~\cite{Guanbo} We investigate its finite-sample performance in the simulation study. In past work, we have also found that the nonparametric clustered bootstrap performs well.\cite{Mireille2014, Guanbo}

\section{Simulation Study}
\label{Simulation Study}
We conducted simulation studies in R to demonstrate the double robustness and finite-sample performance of the TMLE. Results for the A-IPTW estimator are provided in the Supplementary Materials Appendix E.

\subsection{Data Generation}
\label{Data Generation}
 For each dataset, we generated two continuous study-level covariates, $S_1$ and $S_2$, where we treated $S_2$ as unobserved. We also generated three individual-level covariates: $W_1$ continuous and $W_2$ and $W_3$  binary. In this simulation, we included three treatments where each study had access to one, two or all three. We denote the treatment availability as $D^{(k)},k = 1,2,3$ which was generated conditional on the study level covariate $S_1$. Then three treatment indicators $A^{(k)}$ were generated based on the values of $S_1, W_1, W_2, W_3$ and $D^{(k)}$. Subjects could take any combination of these treatments. Finally, we generated a continuous outcome $Y$ in a model with and without random effects, conditional on individual-level covariates, all three treatments, and study-level covariate $S_1$. Specifically, this model included interactions between the two effect modifiers $W_1$ and $W_3$ and $A^{(1)}$. For random effects by study we added an additional interaction term between $S_2$ and $A^{(1)}$. Table S1 in Supplementary Material Appendix E displays the full data generating mechanism. The observed data structure for each subject is $O = (S_1,W_1, W_2, W_3,D,A,Y)$.

In this simulation study, we only aimed to estimate the parameters representing effect modification of treatment $k=1$. For both scenarios with the outcomes involving random effects or not, we drew 1000 simulations  with $J\in \{10,30,50\}$ studies where each study contained 300 subjects. This resulted in three total sample sizes, $n=3000,9000,15000$. 

\subsection{Analysis}
\label{Analysis}
In order to model the effect modification of treatment $k=1$, we included treatment covariates $A^{(2)}$ and $A^{(3)}$ as confounders, as explained in Section~\ref{Parameter of Interest and Assumptions}. We denote the baseline covariates as $\boldsymbol{X}^{(1)}=\{S_1, W_1,W_2,W_3,A^{(2)}, A^{(3)}\}$. In practice we are not aware of the true set of effect modifiers so we define the potential set as $\{W_1,W_2,W_3,A^{(2)}, A^{(3)}\}$ and set $\boldsymbol{V}^{(1)}=\{1,W_1,W_2,W_3,A^{(2)}, A^{(3)}\}$.

As we discussed in Section \ref{Parameter of Interest and Assumptions}, we model the CATE of treatment $A^{(1)}$ as a linear function of potential effect modifiers:
$\psi\{\boldsymbol{V}^{(1)};\boldsymbol{\beta}_{\boldsymbol{V}^{(1)}}\}=\{\boldsymbol{V}^{(1)}\}^\intercal\boldsymbol{\beta}_{\boldsymbol{V}^{(1)}}=\boldsymbol\beta_0^{(1)} + \boldsymbol\beta_1^{(1)}W_1 + ... + \boldsymbol\beta_5^{(1)}A^{(3)}$. Our parameters of interest are thus $\boldsymbol{\beta}_{\boldsymbol{V}^{(1)}}$. In the scenarios without random effects, the true values of these parameters were derived analytically. For the scenarios with random effects, we generated data with large sample sizes ($10^6$) and forced all $A^{(1)}$ equal to 1 and 0, respectively, to obtain both counterfactual outcomes. Then we obtained the true values of the parameters by fitting the linear regression of the differences between the two counterfactual outcomes on the potential effect modifiers.  Given the data generating mechanism there were two real effect modifiers, $W_1$ and $W_3$. When outcomes were simulated without random effects, the true corresponding coefficient values were $0.65$ and $0.35$, respectively. When outcomes were generated with random effects, the true values were $0.77$ and $0.38$ (Tables S2, S3 in the Supplementary Materials Appendix E).

 We used logistic regressions to estimate each component of $g\{A|\boldsymbol{X}^{(1)}\}$, collectively referred to as $g$, and $\bar{Q}\{a,\boldsymbol {X}^{(1)}\}$, referred to as $Q$. To control the potential sources of sparsity, predicted values for $g_1$ and $g_2$ were truncated at $(\alpha,1-\alpha)$ where $\alpha=0.001$. To demonstrate the double robustness property of both methods, we ran four different scenarios: 1) correctly specified parametric models for both $Q$ and $g$; 2) only the $Q$ model correctly specified and the $g$ model misspecified as a null model; 3) only the $g$ model correctly specified and the $Q$ model misspecified as a null model; 4) both $Q$ and $g$ misspecified as null models.

 In this simulation, we applied the TMLE algorithms presented in Section \ref{TMLE}. Note that the TMLE implementation we used requires that we transform the continuous Y to lie in $(0,1)$, and then we reverse-transform at the end of the procedure.\cite{van2010} The standard errors were estimated by the influence function sandwich estimator, first ignoring clustering and then incorporating clustering as described in Section~\ref{Influence Function-based Variance Estimator}. We compared both standard error estimates to the Monte Carlo standard errors. Then based on the clustered standard errors, we constructed $95\%$ Wald-type confidence intervals and computed the corresponding coverage rates which are the percentage of times that the confidence intervals contained the true parameter values.

\subsection{Results}
\label{Results}
\begin{figure}[htbp]
 \centering
 \includegraphics[width=16cm,height=18cm]{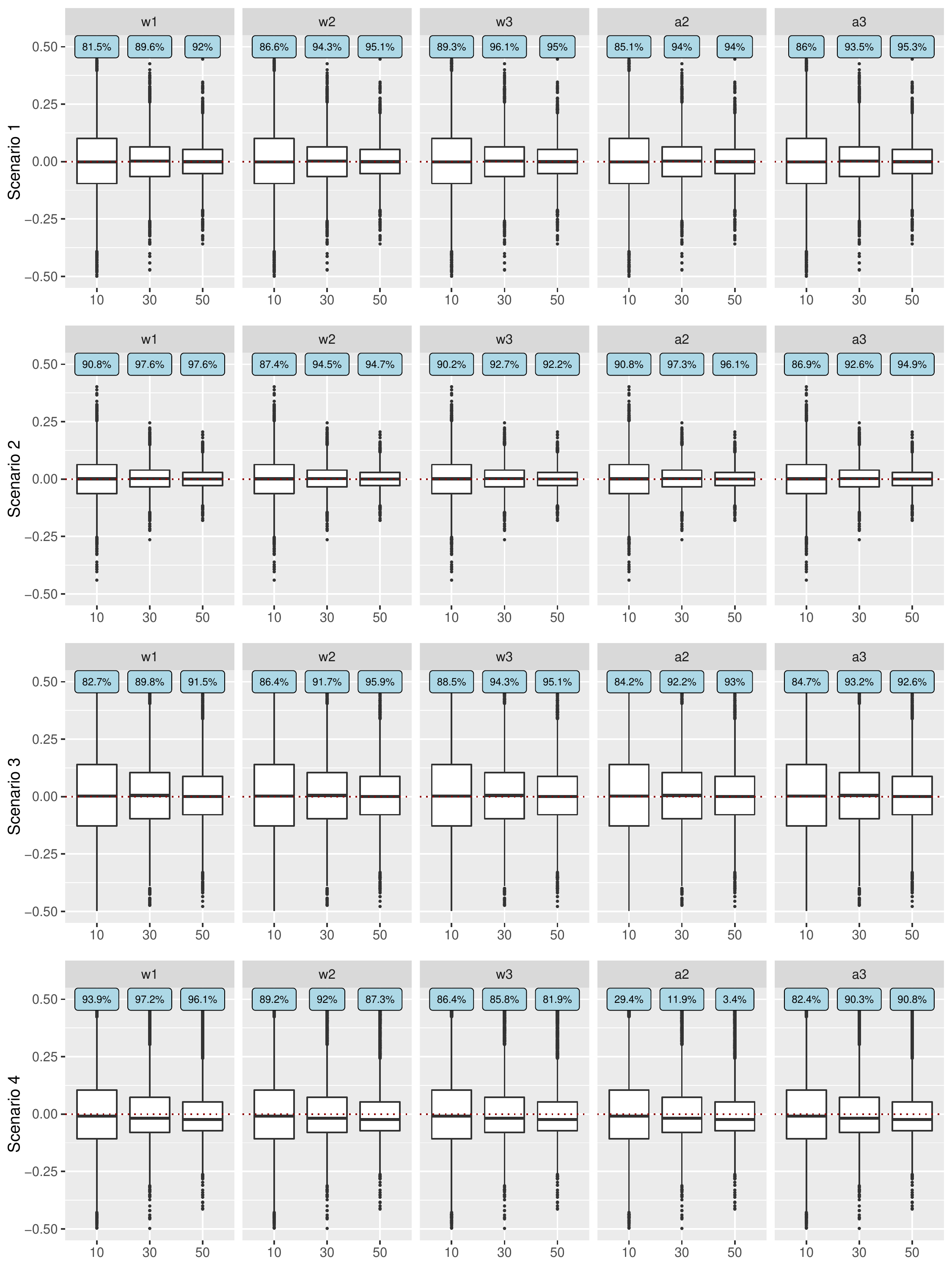}
 \caption{Error of TMLE estimates under four scenarios and three different sample sizes without random effects. The $x$-axis represents the number of studies. Coverage rates based on the clustered sandwich estimators of the standard error are presented in blue boxes. The four scenarios are as follows: Scenario 1 - both $Q$ and $g$ models are correct; Scenario 2 - $Q$ model is correct, $g$ is null; Scenario 3 - $Q$ model is null, $g$ model is correct; Scenario 4 - both $Q$ and $g$ models are null.}
 \label{fig:simu_noRE}
 \end{figure}

\begin{figure}[htbp]
 \centering
 \includegraphics[width=16cm,height=18cm]{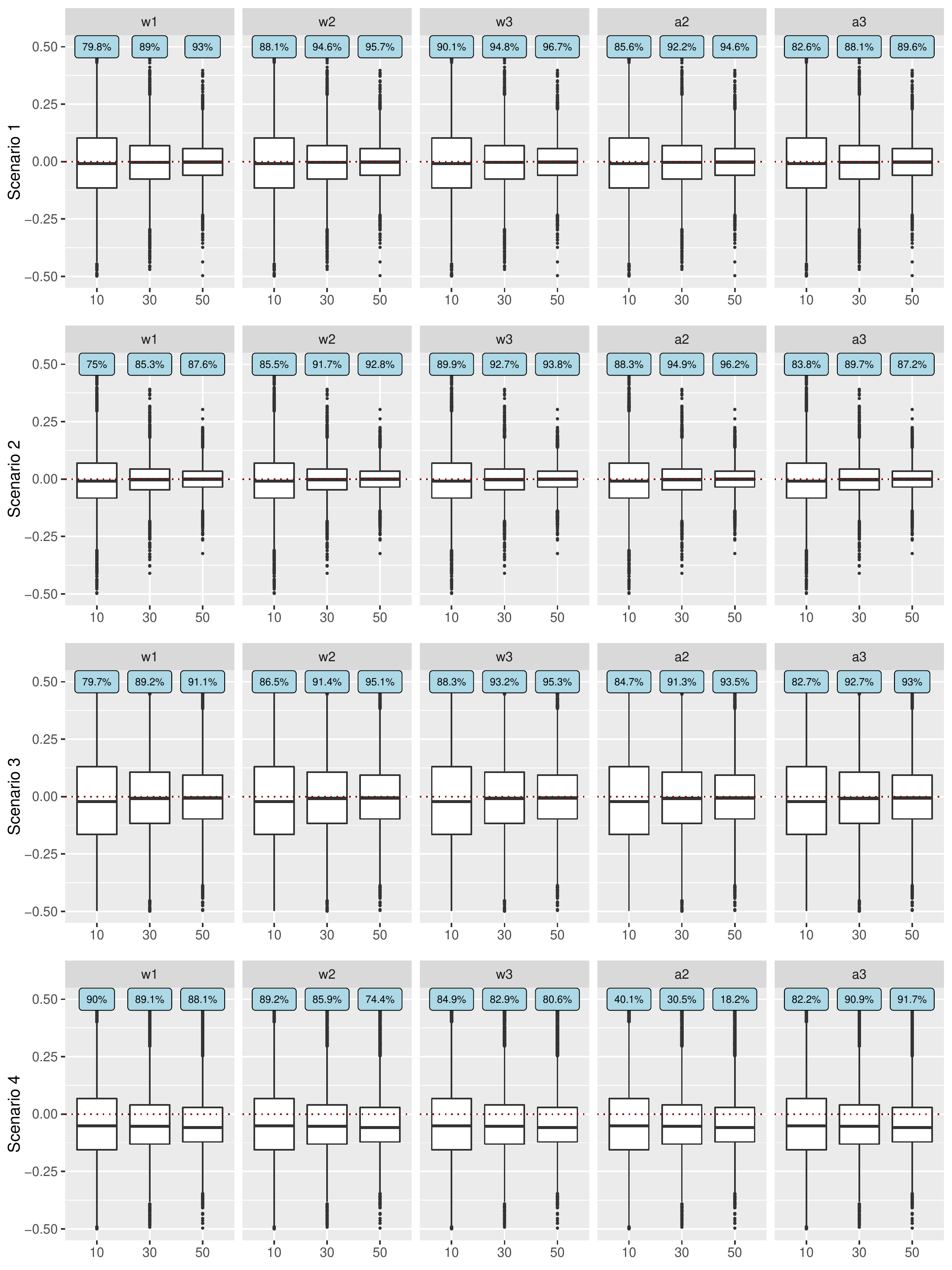}
 \caption{Error of TMLE estimates under four scenarios and three different sample sizes with random effects. The $x$-axis represents the number of studies for three sample sizes. Coverage rates based on the clustered sandwich estimators of the standard error are presented in blue boxes. The four scenarios are as follows: Scenario 1 - both $Q$ and $g$ models are correct; Scenario 2 - $Q$ model is correct, $g$ is null; Scenario 3 - $Q$ model is null, $g$ model is correct; Scenario 4 - both $Q$ and $g$ models are null.}
 \label{fig:simu_RE}
 \end{figure}

 Figure \ref{fig:simu_noRE} presents the results of TMLE for the estimation of five potential effect modifiers under the four estimation scenarios and three different sample sizes with outcomes generated without random effects. Figure \ref{fig:simu_RE} presents the TMLE results under random effects. The coverage rates are presented in the blue boxes of the two figures. More detailed TMLE results are provided in the Supplementary Materials Appendix E (Tables S2 - S4), along with the figures and tables of the results of the A-IPTW estimator (Tables S5 - S7 and Figures S1, S2). 
 
 From Figures \ref{fig:simu_noRE} and \ref{fig:simu_RE}, we see that under the first two scenarios, where the model for $Q$ was correctly specified, the TMLE estimators had no error on average regardless of the presence of random effects. Unsurprisingly, under scenario 4 when all quantities were assigned null models, a small bias was present without random effects and a larger bias was present with random effects.  In scenario 3, where the model for $Q$ was incorrectly specified but the model for $g$ was correct, the average error converged to zero as the number of studies increased. The estimates were also more dispersed than for the previous scenarios. This occurred because the model for $g_2$ is estimated using the study as the unit in the analysis, so with few studies, the sample size to fit this model is extremely small. Compared to TMLE, the A-IPTW estimator had greater average error in scenarios 3 and 4, but otherwise performed similarly.
 
 The coverage rates, given in the blue boxes of Figures \ref{fig:simu_noRE} and \ref{fig:simu_RE}, typically increased with the number of studies. For instance, with random effects with 10 studies, in scenario 1 where all models were correctly specified, the coverage rates for the five potential effect modifiers were between $79.8\%$ and $90.1\%$. With 30 studies, the rates increased to  $88.1\%-94.8\%$. Then for 50 studies, the coverage rates were $89.6\%-96.7\%$. Similar patterns were also obtained in the second and third scenarios. 
 
\section{MDR-TB Data Analysis}
\label{MDR-TB Data Analysis}
\subsection{Descriptive Statistics of MDR-TB }
\label{Descriptive Statistics of MDR-TB}
The combined dataset contained the IPD from 31 observational studies with a total of 9290 patients. After removing 260 ($2.8\%$) patients without reported outcomes, 9030 patients taking 15 different antimicrobial agents remained. Study-specific sample sizes ranged from 25 to 2182 patients. Missing values were present in the covariates; the cavity status variable had the most missing values ($25.97\%$). 
Appendix F Table S8 and Figure S3 give summaries of the treatment-specific sample sizes and the six individual level covariates. Ofloxacin, pyrazinamide and cycloserine were the three most prescribed medications. In contrast, fewer than 1000 patients were prescribed amikacin, later-generation fluoroquinolones and ciprofloxacin. Ethambutol and pyrazinamide were widely prescribed in 30 out of 31 studies, while only 14 studies had patients who took ciprofloxacin or later-generation fluoroquinolones. The number of male patients was around twice  the number of female patients in all treatment groups. A total of 4892 ($54.33\%$) patients were concentrated in the $26-45$ year-old age group while only 154 ($1.71\%$) were in the $0-17$ year-old group. For the other individual level covariates, about $75\%$ of patients were diagnosed with TB in the past. Moreover, the majority of patients had cavity ($68.33\%$) and positive AFB ($74.53\%$), but there were only $15\%$ coinfected with HIV. 

In order to assess the data support needed to investigate effect modification by concurrent medication, Table \ref{table:MDR-TB} displays the number of patients who used a combination of any two medications, with the diagonal indicating the total number of patients taking the corresponding medication. Values ranged between 58 to 4574, indicating at least minimal data support for all pairwise combinations. It should be noted that the $558$ patients prescribed amikacin all took kanamycin during the study period. This indicates a positivity violation for the analysis of the effect of kanamycin if we were to treat amikacin as a potential confounder of kanamycin. It also means that we cannot evaluate the effect modification of amikacin by kanamycin. Therefore, we pragmatically excluded amikacin both as a confounder and as an effect modifier in the analysis of kanamycin. And in the analysis of amikacin, kanamycin was excluded as a potential effect modifier but was still included as a confounder. 

%\begin{landscape}
\begin{table}[!ht]
\small\centering
\caption{Summary of number of patients taking any combinations for any two medications during the treatment period. The diagonal values represent the total number of patients taking each medication.}
\label{table:MDR-TB}
\resizebox{\columnwidth}{!}{%
\begin{tabular}{rrrrrrrrrrrrrrrr}
  \toprule
 & \textbf{EMB} & \textbf{AMK} & \textbf{CAP} & \textbf{CIP} & \textbf{CS} & \textbf{ETO} & \textbf{OFX} & \textbf{PAS} & \textbf{PTO} & \textbf{RIF} & \textbf{SM} & \textbf{PZA} & \textbf{KMA} & \textbf{LgFQ} & \textbf{Gp5} \\ 
  \midrule
\textbf{EMB} & \textbf{4188} & 266 & 734 & 565 & 1617 & 2472 & 2995 & 989 & 770 & 1080 & 634 & 3518 & 2706 & 272 & 768 \\
  \textbf{AMK} & 266 & \textbf{558} & 181 & 129 & 416 & 167 & 341 & 271 & 187 & 154 & 58 & 340 & 558 & 128 & 339 \\ 
  \textbf{CAP} & 734 & 181 & \textbf{1874} & 482 & 1719 & 900 & 1328 & 1386 & 722 & 325 & 151 & 1119 & 543 & 205 & 923 \\ 
  \textbf{CIP} & 565 & 129 & 482 & \textbf{968} & 782 & 682 & 236 & 616 & 223 & 350 & 232 & 645 & 481 & 127 & 555 \\
  \textbf{CS} & 1617 & 416 & 1719 & 782 & \textbf{5629} & 1941 & 4195 & 3573 & 3004 & 523 & 981 & 3234 & 2900 & 745 & 1833 \\
  \textbf{ETO} & 2472 & 167 & 900 & 682 & 1941 & \textbf{3911} & 3175 & 1206 & 240 & 397 & 309 & 3433 & 3014 & 287 & 670 \\
  \textbf{OFX} & 2995 & 341 & 1328 & 236 & 4195 & 3175 & \textbf{6464} & 2750 & 2566 & 465 & 786 & 4574 & 4191 & 192 & 1262 \\ 
  \textbf{PAS} & 989 & 271 & 1386 & 616 & 3573 & 1206 & 2750 & \textbf{3937} & 2292 & 293 & 732 & 1962 & 1816 & 644 & 1463 \\ 
  \textbf{PTO} & 770 & 187 & 722 & 223 & 3004 & 240 & 2566 & 2292 & \textbf{3304} & 154 & 749 & 1564 & 1532 & 449 & 1065 \\
  \textbf{RIF} & 1080 & 154 & 325 & 350 & 523 & 397 & 465 & 293 & 154 & \textbf{1261} & 406 & 1133 & 332 & 87 & 195 \\ 
  \textbf{SM} & 634 & 58 & 151 & 232 & 981 & 309 & 786 & 732 & 749 & 406 & \textbf{1366} & 870 & 192 & 269 & 339 \\ 
  \textbf{PZA} & 3518 & 340 & 1119 & 645 & 3234 & 3433 & 4574 & 1962 & 1564 & 1133 & 870 & \textbf{6102} & 3775 & 436 & 930 \\
  \textbf{KMA} & 2706 & 558 & 543 & 481 & 2900 & 3014 & 4191 & 1816 & 1532 & 332 & 192 & 3775 & \textbf{5015} & 416 & 1166 \\
  \textbf{LgFQ} & 272 & 128 & 205 & 127 & 745 & 287 & 192 & 644 & 449 & 87 & 269 & 436 & 416 & \textbf{866} & 511 \\
  \textbf{Gp5} & 768 & 339 & 923 & 555 & 1833 & 670 & 1262 & 1463 & 1065 & 195 & 339 & 930 & 1166 & 511 & \textbf{2138} \\ 
   \bottomrule
\end{tabular}
}
\end{table}
%\end{landscape}

\subsection{Analysis and Results of MDR-TB}
\label{Analysis and Results of MDR-TB}
For each medication $k$, the target parameters in this application are the coefficients of the MSMs, $$\psi\{\boldsymbol{V}^{(k)};\boldsymbol{\beta}_{\boldsymbol{V}^{(k)}}\}=\{\boldsymbol{V}^{(k)}\}^\intercal\boldsymbol{\beta}_{\boldsymbol{V}^{(k)}}=\beta_0^{(k)}+\sum_{j=1}^{20}V^{(k)}_j\beta^{(k)}_j$$
where $\boldsymbol{V}^{(k)}=\{1,V_1^{(k)},\cdots,V_{20}^{(k)}\}$ is the set including six individual level covariates and the 14 medications excluding medication $k$. We standardized the continuous variable age. $\beta_0^{(k)}$ represents the baseline effect for the reference group of female patients with mean age (39), negative AFB test, no cavitation on chest radiography, no past TB history, no HIV co-infection and not taking any other of the 14 medications. The associated coefficients $\beta^{(k)}_j,j=1,\cdots,20$ represent the difference in the CATE when varying the characteristic $V^{(k)}_j$ by one unit while holding other covariates fixed.

As noted, there are missing values in the individual-level covariates. We used multiple imputation \cite{MI} by chained equations with the \texttt{MICE} package~\cite{MICE} in \texttt{R} to produce $20$ imputations then used Rubin's rules to combine the estimates.\cite{Rubin's rule} In each imputed dataset we followed the procedure described in Section~\ref{TMLE} to fit a TMLE. We estimated the $Q$ component using SuperLearner (SL)~\cite{van_SL} which is a methodology that uses cross validation to find an optimal convex combination of the predictions of a library of candidate algorithms defined by the user. We included the following algorithms in the SL library: generalized linear models with penalized maximum likelihood (\texttt{glmnet} function)~\cite{glmnet, TibshiraniLASSO}, with forward stepwise variable selection  (\texttt{step} function), and with a stepwise procedure based on the Akaike Information Criterion (\texttt{stepAIC} function), respectively.~\cite{step}  Logistic regressions were used for the $g$ models and LASSO penalties were added when the logistic regression failed to converge. 

 In each completed dataset, variance estimates for the coefficients were computed using the sample variance of the influence function following the expression in Section~\ref{Influence Function-based Variance Estimator}. Finally, since many comparisons made in this analysis, we performed a multiple testing adjustment of the ``significance level'' of the p-values to control the false discovery rate via the method of Benjamini and Hochberg ~\cite{fdr} with further details given in the Supplementary Materials Appendix G. 

\begin{figure}[!htbp]
 \centering
 \includegraphics[width=16cm,height=18cm]{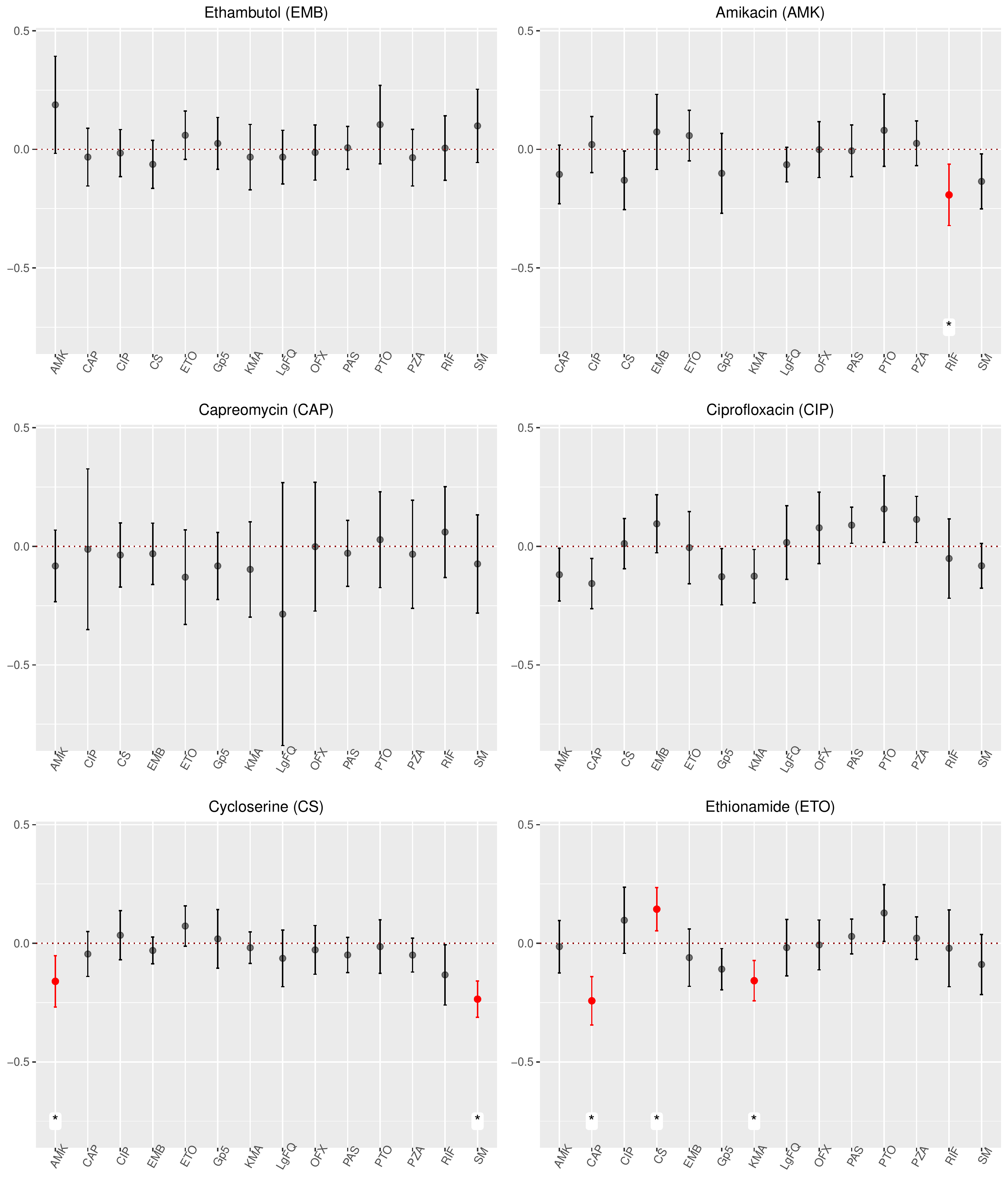}
 \caption{Estimated coefficients of potential effect modifications and the corresponding $95\%$ confidence intervals for EMB, AMK, CAP, CIP, CS and ETO. Significant results are shown in red and indicated with $\ast$.}
 \label{fig:res_com1}
 \end{figure}

\begin{figure}[!htbp]
 \centering
 \includegraphics[width=16cm,height=18cm]{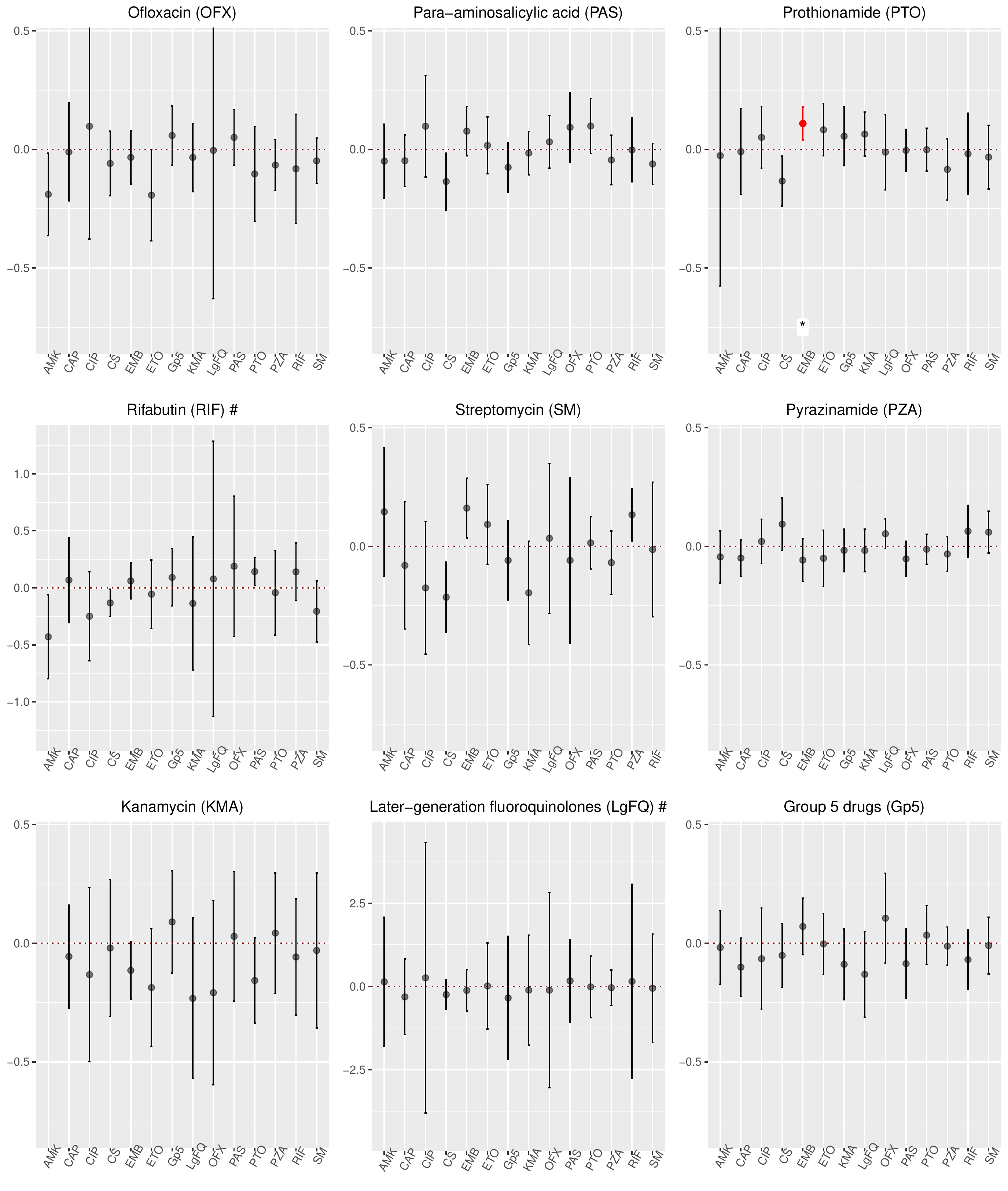}
  \caption{Estimated coefficients of potential effect modifications and the corresponding $95\%$ confidence intervals for OFX, PAS, PTO, RIF, SM, PZA, KMA, LgFQ and Gp5. Significant results are shown in red and indicated with $\ast$.\\
 \# Large scale in $y$-axis of RIF plot and LgFQ plot. } 
 \label{fig:res_com2}
 \end{figure}

Figures \ref{fig:res_cov}, \ref{fig:res_com1} and \ref{fig:res_com2} show the estimated coefficients, standard errors and $95\%$ confidence intervals corresponding to the intercept and six individual-level effect modifiers and the other 14 medications. The outputs for estimated coefficients that are statistically significant after adjustment are marked in red color and asterisk. Corresponding tables of the numerical results are also provided in the Supplementary Materials Appendix F (Tables S9 and S10). From Figure~\ref{fig:res_cov} there is no evidence that the six characteristics modify the treatment effect of any drug. In Figure \ref{fig:res_com1} (AMK plot), we see that patients prescribed rifabutin had lower estimated effects of amikacin. Patients prescribed streptomycin or amikacin might have benefited less from cycloserine than patients not taking streptomycin or amikacin (CS plot). In addition, taking cycloserine was associated with a greater estimated effect of ethionamide while capreomycin, kanamycin drugs were associated with lower estimated effects (ETO plot). Finally, in Figure \ref{fig:res_com2} (PTO plot), we interpret that patients prescribed ethambutol may have benefited more from prothionamide than patients not taking ethambutol.

The empirical distributions of the untruncated propensity scores for all drugs are provided in the Supplementary Materials Appendix F (Table S11 and Figure S4). For later-generation fluoroquinolones, we noted very large weights which likely yielded the large variability observed in Figure \ref{fig:res_com2}, LgFQ plot. In addition, since fewer patients were prescribed later-generation fluoroquinolones among those who were HIV positive (Appendix F Figure S3), the standard errors were inflated for the coefficient of HIV in the fluoroquinolones MSM (Figure \ref{fig:res_cov}, LgFQ plot). Also, among those who took rifabutin, only 87/866 subjects were also prescribed later-generation fluoroquinolones, giving rise to the large standard error (Figure \ref{fig:res_com2}, RIF plot). 

\begin{figure}[!htbp]
 \centering
 \includegraphics[width=16cm,height=18cm]{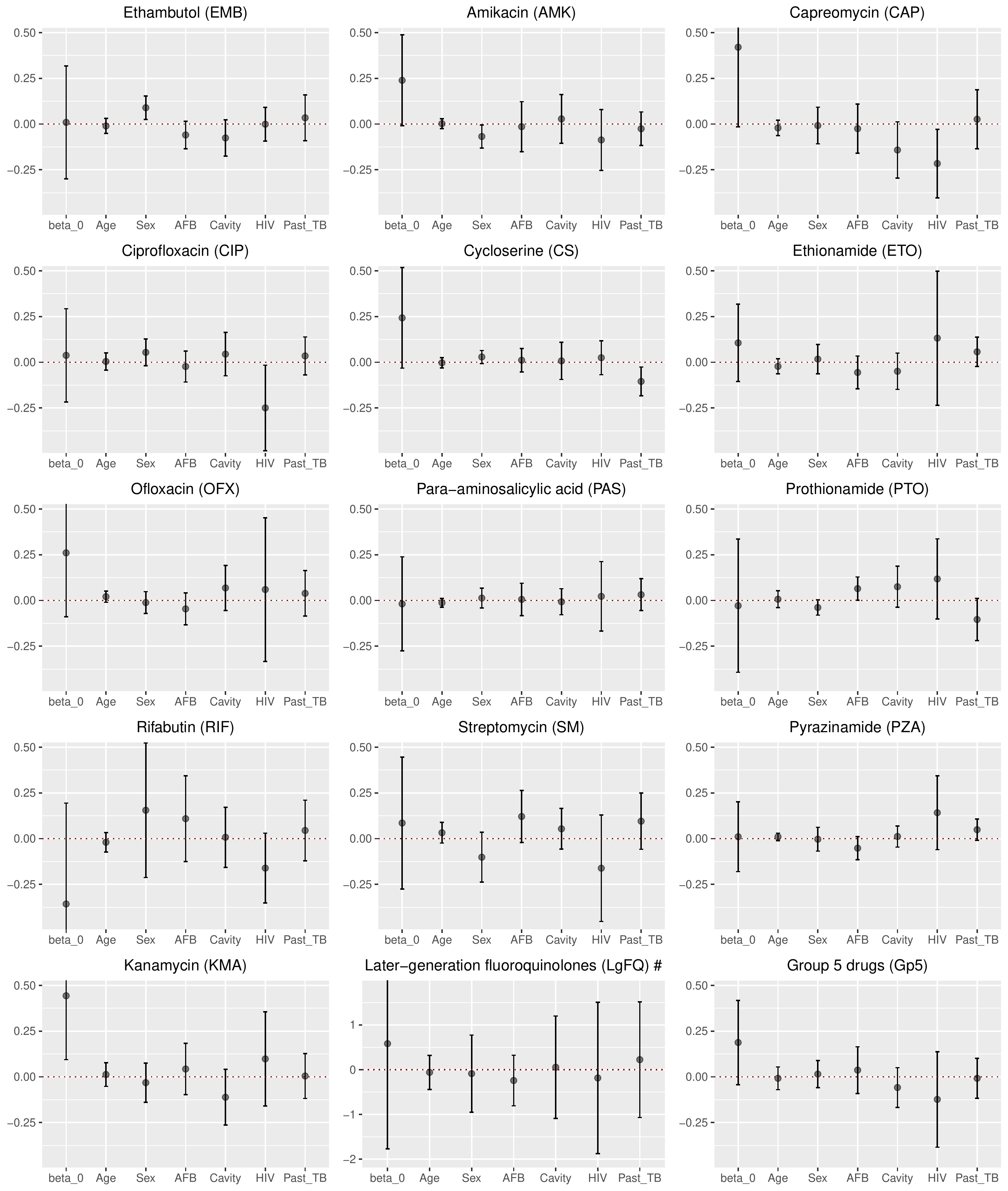}
  \vspace{0.4cm}
 \caption{Estimated coefficients and the corresponding $95\%$ confidence interval for 15 medications relative to the intercept and six covariates. None of the coefficients reached statistical significance.\\
 \# Large scale in $y$-axis of LgFQ plot. }
 \label{fig:res_cov}
 \end{figure}
 
\section{Discussion}
\label{Discussion}
In this paper, we developed two related one-stage doubly-robust approaches to the analysis of baseline effect modification in an IPD meta-analysis.
The model space that we considered was nonparametric though the parameters of interest were defined through a working linear MSM for the CATE. These approaches allowed us to analyze pooled IPD from multiple studies in order to evaluate how estimated treatment effects may vary depending on the values of patient covariates. Our past work, which proposed related methods and a TMLE for IPD meta-analysis with multiple treatments, instead estimated a treatment importance metric defined as the difference in adjusted probabilities of treatment success between the patients who used each medication and the overall population.~\cite{Guanbo} 

Vo et al. illustrated that in IPD meta-analysis, heterogeneity across studies can come from two sources: case-mix heterogeneity, due to effect modification, and beyond case-mix heterogeneity, due to differences in study design and measurement.\cite{tat2019} Our methods address this by allowing for differential availability of treatments across studies and random effects by study due to measured and unmeasured characteristics of the study-specific populations.

In clinical and epidemiological research, model misspecification is always a concern when estimating treatment or exposure effects. Doubly robust methods yield consistent estimators even under misspecification of either the treatment or the outcome model. In the simulation study, we demonstrated the double robustness property of both TMLE and A-IPTW. We observed similar performance of TMLE and A-IPTW but we did not investigate near-positivity violations or other scenarios that may differentiate them in finite samples as have others.~\cite{Porter,Mireille2013} In addition, we demonstrated the double robustness of both methods when there are study-specific random effects for the outcome. Finally, we showed that the proposed confidence intervals, estimated using the influence function sandwich estimator that considers clustering by study, performed well when there were greater than 30 studies in the analysis. Indeed, a limitation of our approach is that it relies on a sufficient number of studies to estimate a generalizable parameter. In particular, the ability to adjust for confounding by treatment availability depends on fitting a model for treatment availability conditional on study-level covariates, which is limited by the typically small number of studies in a meta-analysis. Indeed, we observe in the simulation study that error may persist when the number of studies is small and the outcome regression model is incorrectly specified, even when the propensity score model components are correctly specified with parametric models. Therefore, our approach should only be undertaken when a larger number of studies are available.

The treatment of MDR-TB has been challenging because of its prolonged duration, toxicity, costs and unsatisfactory outcomes.~\cite{WHO} Second-line TB medicines used for the treatment of drug-resistant TB include amikacin, ciprofloxacin, cycloserine, capreomycin, ethionamide, later-generation fluoroquinolones, kanamycin, ofloxacin, para-aminosalicyclic acid, prothionamide and streptomycin.  
We used TMLE to investigate effect modification of each of the 15 observed medications by other individual-level covariates, including the medications taken concurrently for the treatment of MDR-TB. Our results suggest that 
cycloserine may enhance the effects of ethionamide but capreomycin and kanamycin were associated with reduced effects. In addition, amikacin and streptomycin may reduce the effect of cycloserine and ethambutol might boost the effect of prothionamide. For the six individual characteristics, age, sex, acid fast bacilli status, HIV infection, cavitation status on chest radiography, and past TB history, we did not find any evidence of effect modification. However, since our MDR-TB data were identified from studies carried out up to 2009, we have no information about both new and repurposed effective anti-TB drugs.~\cite{Pym,Sotgiu} Therefore, it would be expected to apply our methods to data from more recently treated patients.

Another limitation of our approach is that, because we only considered the effect of intervening on one treatment at a time, we cannot directly address how to select combinations of medications that would be expected to optimize the probability of treatment success. Previous work evaluated the causal contrasts between different regimens of concurrent medications in MDR-TB.~\cite{siddique} Future applications should directly address the more challenging question of treatment-treatment interactions on the outcome which would directly allow for the evaluation of optimal medication usage. Other ongoing work in our group involves using LASSO,~\cite{TibshiraniLASSO} rather than hypothesis testing, to select the effect modifiers in the linear MSM for the CATE. This may improve upon the current work by utilizing a superior approach to variable selection.

Identifying effect modifiers is an important step for estimating subpopulation causal effects that can help guide treatment decision making for individual patients. Our findings suggest a way to explore the data and generate hypotheses for drug combinations which can be tested in randomized controlled trial. However, such analyses require larger amounts of data than the estimation of average treatment effects. We have contributed by extending existing doubly robust methods to incorporate multiple data sources. Advances in IPD meta-analysis enable researchers to incorporate multiple sources of previously collected observational data in their analyses in order to increase their power, which is greatly beneficial for the identification of effect modifiers.

\end{document}